\documentclass[letterpaper]{ptephy_v1}

\usepackage{url} 

\newcommand{\msun}{{\rm M}_\odot}

\begin{document}

\title{Formation of mass gap compact object and black hole binary  from Population III stars}

\author[1]{Tomoya Kinugawa}
\author[2]{Takashi Nakamura}
\author[3]{Hiroyuki Nakano}

\affil[1]{Institute for Cosmic Ray Research, The University of
  Tokyo, Kashiwa, Chiba 277-8582, Japan}

\affil[2]{Department of Physics, Graduate School of Science, Kyoto University, Kyoto 606-8502, Japan}

\affil[3]{Faculty of Law, Ryukoku University, 
Kyoto 612-8577, Japan}

\date{\today}

\begin{abstract}%
 We performed population synthesis simulations of Population III binary stars with Maxwellian kick velocity distribution when MGCOs (Mass Gap Compact Objects with mass 2--5$\,M_{\odot}$) are formed. We found that for eight kick velocity dispersion models of $\sigma_{\rm k}=0$--$500$  km/s, the mean mass of black hole (BH)-MGCO binary is $\sim (30 \,M_\odot,\,2.6 \,M_\odot)$. In numerical data of our simulations, we found the existence of BH-MGCO binary with mass $(22.9 \,M_\odot,\,2.5 \,M_\odot)$ which looks like GW190814.
\end{abstract}

\subjectindex{xxxx, xxx}

\maketitle

\section{Introduction}

GW190814~\cite{Abbott:2020khf} is
a gravitational wave (GW) event
observed by LIGO-Virgo collaboration during 
the first half of the third observation period,
called O3a.
This is a compact binary coalescence and the parameter estimation suggests that the binary consists of
a black hole (BH) with mass of 22.2--24.3$\,M_{\odot}$
and a compact object with mass of 2.50--2.67$\,M_{\odot}$ yielding a chirp mass of 
$M_{\rm chirp}\sim 6.03$--$6.15\,M_{\odot}$~\footnote{
 $M_{\rm chirp} = (m_1 m_2)^{3/5}/(m_1+m_2)^{1/5}$ where $m_1$ and $m_2$ denote the mass of the primary and the secondary objects, respectively.}.
The mass of the smaller (secondary) object lies in
2--5$\,M_{\odot}$, that is,
between known neutron stars (NSs) and BHs.
Therefore, the secondary compact object
will be a NS with the maximum observed mass
or a BH with the minimum observed mass,
so that we define a mass gap compact object (MGCO)~\footnote{MGCO is different from ``mass gap'' (2.5--5$\,M_{\odot}$)
used in Ref.~\cite{Abbott:2020khf}
and ``MassGap'' (3--5$\,M_{\odot}$) defined
in Ref.~\cite{glossary}.
MGCOs include not only MassGap BHs,
but also NSs with mass $\gtrsim 2\,M_{\odot}$,
i.e., the mass of MGCOs lies in 2--5$\,M_{\odot}$.}
 as a compact object having mass 2--5$\,M_{\odot}$.
The merger rate density of this type of binaries
is estimated as 1--23$\,{\rm Gpc}^{-3}{\rm yr}^{-1}$~\cite{Abbott:2020khf}.
So far, many proposals and discussions on
the BH-MGCO binary with MGCO mass of $\sim 2.6\,M_{\odot}$ 
were appeared after the announcement of
GW190814 (see, e.g., Refs.~\cite{Rastello:2020sru,Broadhurst:2020cvm,Zevin:2020gma,
Most:2020bba,Vattis:2020iuz,Tan:2020ics,Safarzadeh:2020ntc,
Essick:2020ghc,Zhang:2020zsc,Jedamzik:2020omx,Tsai:2020hpi,
Fattoyev:2020cws,Mandel:2020cig,Yang:2020xyi,
Tsokaros:2020hli,Tews:2020ylw,Clesse:2020ghq,Lim:2020zvx,
Dexheimer:2020rlp,Sedrakian:2020kbi,Roupas:2020jyv,
Godzieba:2020tjn}).

We discussed in our previous study~\cite{Kinugawa:2016skw} in 2016, 
the detection rate and the chirp mass distribution of
NS-BH binaries with mass of NS below $3\,M_{\odot}$
by the population synthesis simulations of
Population III (Pop III) stars.
We found that the merger rate density of Pop III NS-BH binaries is $\sim 1\,{\rm Gpc}^{-3}{\rm yr}^{-1}$
although it depends on the natal kick velocity of NSs. 
We also found that the chirp mass distribution of Pop III
NS-BH binaries
that merge within the Hubble time has a peak
around 6$\,M_{\odot}$.

The results presented in Ref.~\cite{Kinugawa:2016skw}
seems to be consistent with the estimation of Ref.~\cite{Abbott:2020khf}
since in Ref.~\cite{Kinugawa:2016skw} we defined NS if its mass is below 3$\,M_{\odot}$ while the mass of secondary of GW190814 is 2.50--2.67$\,M_{\odot}$ and the chirp mass of GW190814 is evaluated as 6.03--6.15$\,M_{\odot}$.
Thus, it is important to revisit our previous paper
using the definition of MGCO
and try to answer the question raised as follows.
What are 
``the processes by which the lightest BHs
or the most massive NSs form'' in Ref.~\cite{Abbott:2020khf} ?

In this letter, we summarize our previous study
with additional analyses, 
and present a scenario to explain the BH-MGCO
binary of GW190814.

\section{Analysis}

In Ref.~\cite{Kinugawa:2016skw},
we have calculated NS-BH formations
and estimated the number of NS-BH binaries
merging within the Hubble time
by using the population synthesis simulations
of Pop III stars~\cite{Kinugawa:2014zha}.
The key ingredient is to introduce
the NS kick velocity of 200--500$\,{\rm km/s}$
that is evaluated from the observation
of the proper motion of the pulsar.
The effect of kick velocity is important to decrease
the merging time of NS-BH binaries.
For comparison, we have calculated
not only the Pop III NS-BH binaries,
but also Pop I and II NS-BH binaries.

To do the above analysis in our population synthesis Monte Carlo simulations, 
we have considered 
six metallicity cases from $Z=0$ (Pop III)
to $Z=Z_{\odot}$ (Pop I)
where $Z_{\odot}$ is the solar metallicity,
the initial mass, mass ratio, separation
and eccentricity distribution functions
as binary initial conditions,
the Roche lobe overflow, the common envelope phase,
the tidal effect, the supernova (SN) effect,
and the gravitational radiation as binary interactions,
and two kick velocity models with
$\sigma_{\rm k}=265\,{\rm km/s}$
and $\sigma_{\rm k}=500\,{\rm km/s}$
where $\sigma_{\rm k}$ is the dispersion
of a Maxwellian distribution for kick velocity.

Next, we briefly summarize the results.
The chirp mass of Pop III NS-BH binaries 
merging within the Hubble time
is heavier than those of Pop I and II ones.
The peak values of chirp mass distributions
is $\sim 6\,M_{\odot}$ in the case of Pop III
while it is $\sim 2\,M_{\odot}$ in the cases of Pop I and II.
These peak values almost do not depend on
the kick velocity values $\sigma_{\rm k}$.

The NS-BH merger rates at the present day depend
on the progenitors and kick velocities
(see Table 3 in Ref.~\cite{Kinugawa:2016skw} for the details).
The sum of the merger rates of Pop I and II
becomes 19.7$\,{\rm Gpc}^{-3}{\rm yr}^{-1}$
and 6.38$\,{\rm Gpc}^{-3}{\rm yr}^{-1}$
for $\sigma_{\rm k}=265\,{\rm km/s}$ and $500\,{\rm km/s}$,
respectively.
For the Pop III case, we have found
the NS-BH merger rates at the present day
as 1.25$\,{\rm Gpc}^{-3}{\rm yr}^{-1}$ and
0.956$\,{\rm Gpc}^{-3}{\rm yr}^{-1}$ for $\sigma_{\rm k}=265\,{\rm km/s}$ and $500\,{\rm km/s}$,
respectively.
In the previous study, we have assumed that
the mass range of neutron stars are $1.44$--$3\,M_{\odot}$.
Thus, GW190814 belongs to the previous
Pop III NS-BH result.

In this letter,
we further focus on compact object binaries
which consist of a BH and a MGCO with mass of $2$--$5\,M_{\odot}$.
We calculate $10^6$ Pop III binary evolutions and the merger rates of BH-MGCO binaries,
using the same setup
of the previous Pop III NS-BH case~\cite{Kinugawa:2016skw}.
 {For simplicity, we use the simple SN
remnant model given in Refs.~\cite{Belczynski:2001uc,Kinugawa:2014zha}
which is consistent with the red supergiant problem within error~\cite{Kochanek2020}.
A low mass progenitor whose CO core mass is small ($M_{\rm CO}\lesssim5\,M_{\odot}$)
experiences a SN and the Fe core remains as the remnant.
On the other hand, a high mass progenitor 
($M_{\rm CO}\gtrsim7.6\,M_{\odot}$)
becomes a direct collapse,
and the whole stellar mass becomes the remnant mass.
An intermediate progenitor
($5\,M_{\odot}\lesssim M_{\rm CO}\lesssim7.6\,M_{\odot}$)
experiences a failed SN and the fall backed remnant mass
linearly increases with increasing progenitor mass.}

Here, we assume that the SN remnants which have a mass
less than 5$M_{\odot}$ experience SN kick,
and we use the Maxwellian distribution for the kick velocity distribution.
We calculate not only kick velocity dispersion models
of the previous paper of $\sigma_{\rm k}=265$ km/s
and 500 km/s,
but also $\sigma_{\rm k}=100,~150,~200,~300$, and $400$ km/s. 

\begin{figure}[!t]
    \centering
    \includegraphics[width=0.8\textwidth]{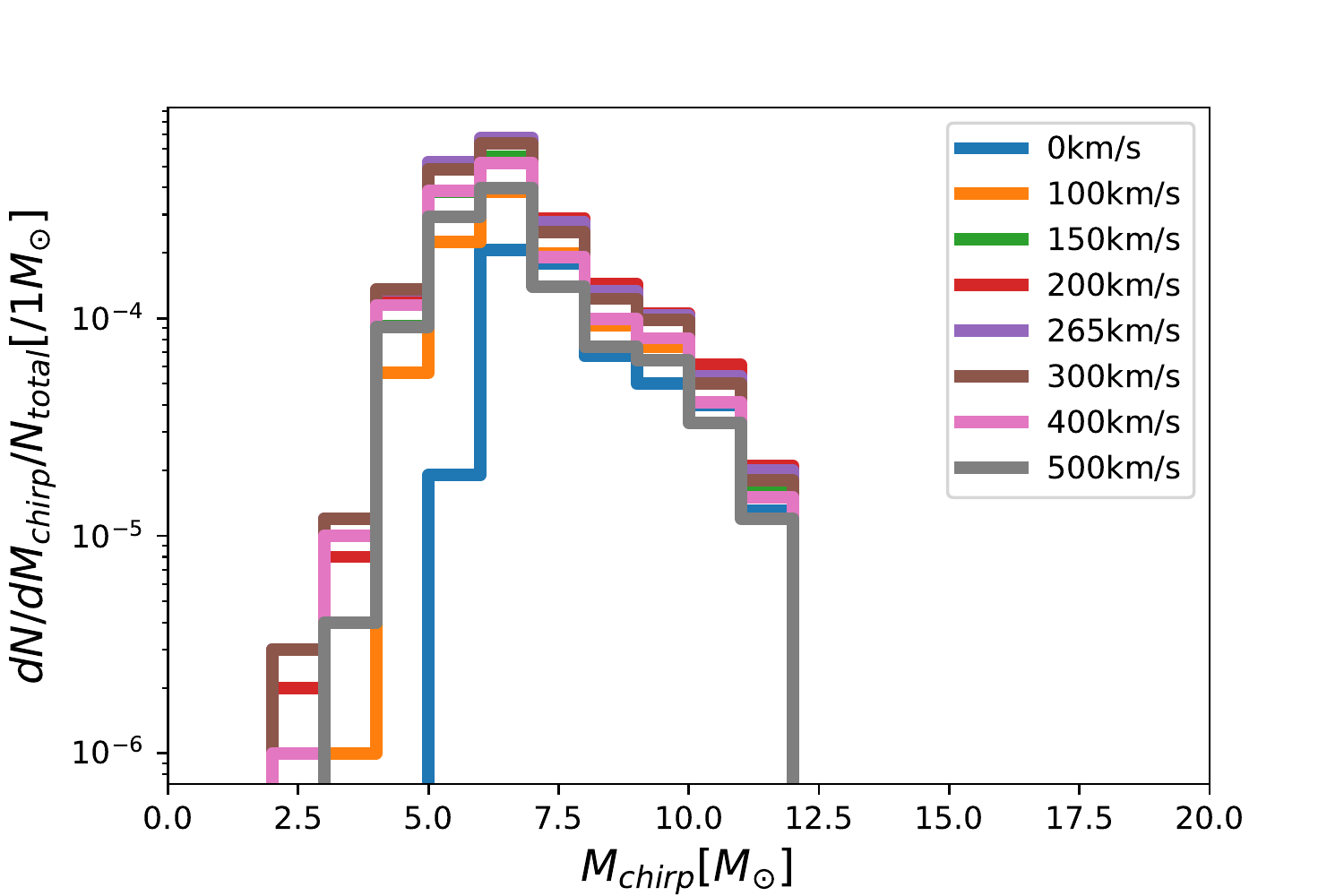}
    \caption{
    The chirp mass distribution of BH-MGCO binaries which merge within the Hubble time for 8 kick models with $\sigma_{\rm k}=0,~100,~150,~200,~265,~300,~400$, and $500$ km/s, where $N_{\rm total}=10^6$ is the total number of binaries
    which we have simulated.
    The chirp mass distribution does not strongly depend on the kick velocity.
    The averaged individual masses are summarized in Table~\ref{tab:ER}.}
    \label{fig:chirpmass}
\end{figure}

\begin{figure}[!t]
    \centering
    \includegraphics[width=0.8\textwidth]{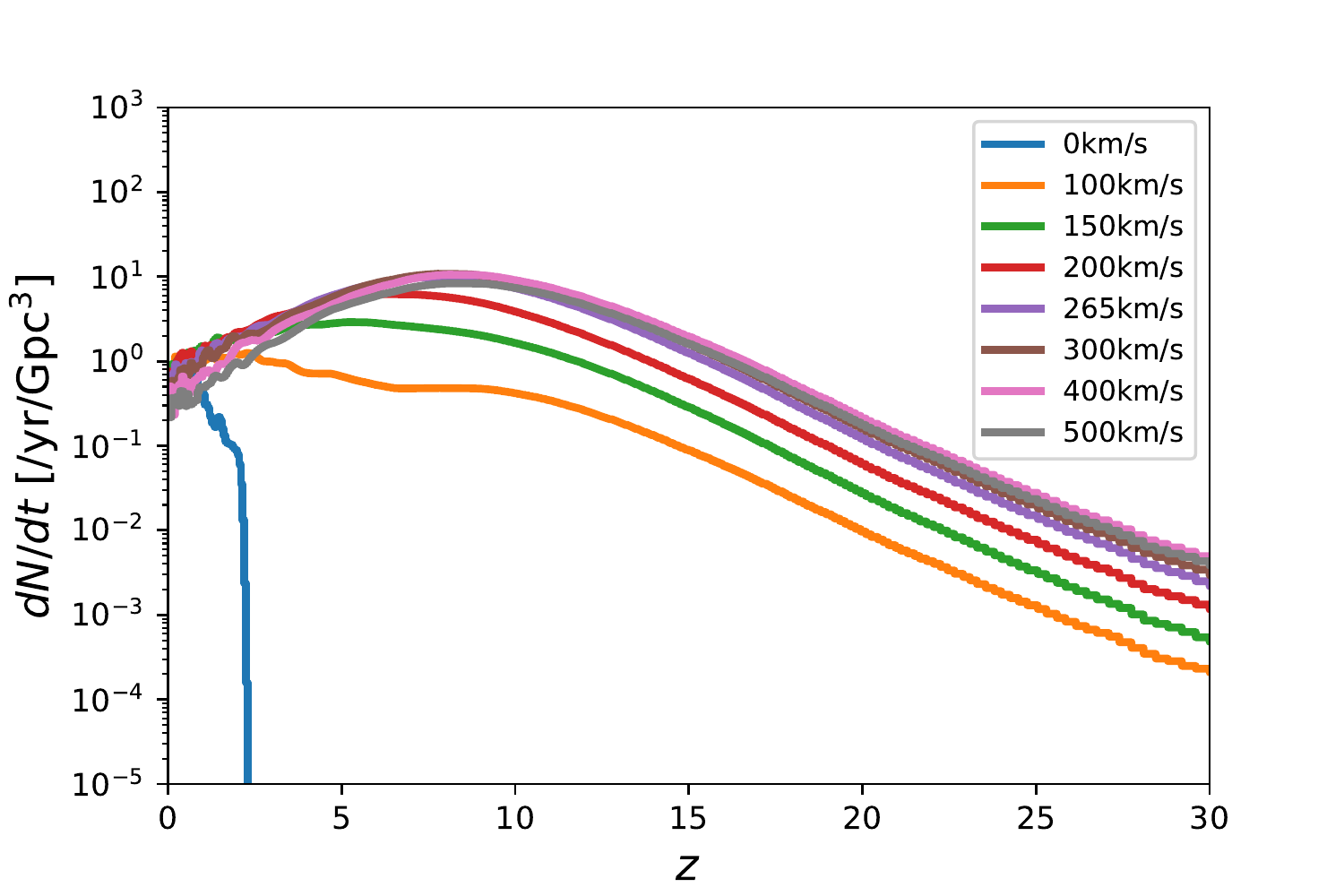}
    \caption{The merger rate densities for 7 kick models with $\sigma_{\rm k}=100,~150,~200,~265,~300,~400$, and $500$ km/s. We note that the smaller kick velocity model, the lower redshift the peak of merger rate is.}
    \label{fig:merger rate}
\end{figure}

Results are shown by Figs.~\ref{fig:chirpmass} and \ref{fig:merger rate}.
Figure \ref{fig:chirpmass} shows the chirp mass distribution of BH-MGCO binaries which merge within the Hubble time for each model, where $N_{\rm total}=10^6$ is the total number of binaries which we have simulated.
The chirp mass distribution does not strongly depend on the kick velocity {and has 
a peak at $M_{\rm chirp} \sim 6 M_{\odot}$. Notice here that the chirp mass of GW190814 is
$\sim 6 M_{\odot}$}.
Figure \ref{fig:merger rate} shows the merger rate densities for each model.
The peak of merger rate density depends on the kick velocity.
Almost all BH-MGCO binaries cannot merge within the Hubble time without the natal kick.
The natal kick changes the binary orbit and makes the binary be able to merge within the Hubble time.
 {If the kick velocity is smaller than the orbital velocity of the BH-MGCO progenitor ($\sim$ a few hundred km/s), the kick only affects as a perturbation like 100 km/s model, and the merger time of BH-MGCO tends to be still long (i.e., they merge at low redshift).
However, if the kick velocity becomes more than the orbital velocity of the BH+MGCO progenitor, the orbit of BH-MGCO can be determined by the kick.
In this case, the kick sometimes brakes the orbit and the orbit of BH-MGCOs becomes so close that they can merge with a short merger time.
Thus, the peak of merger rate density asymptotically moves to near the peak of Pop III SFR ($z\sim10$).
Furthermore, the number of BH+MGCO binaries decreases with the increasing $\sigma_k$.
The merger rate slightly decreases at almost all red shift values if the kick is $\gtrsim 300$ km/s.}

\begin{figure}
    \centering
    \includegraphics[width=1.0\textwidth]{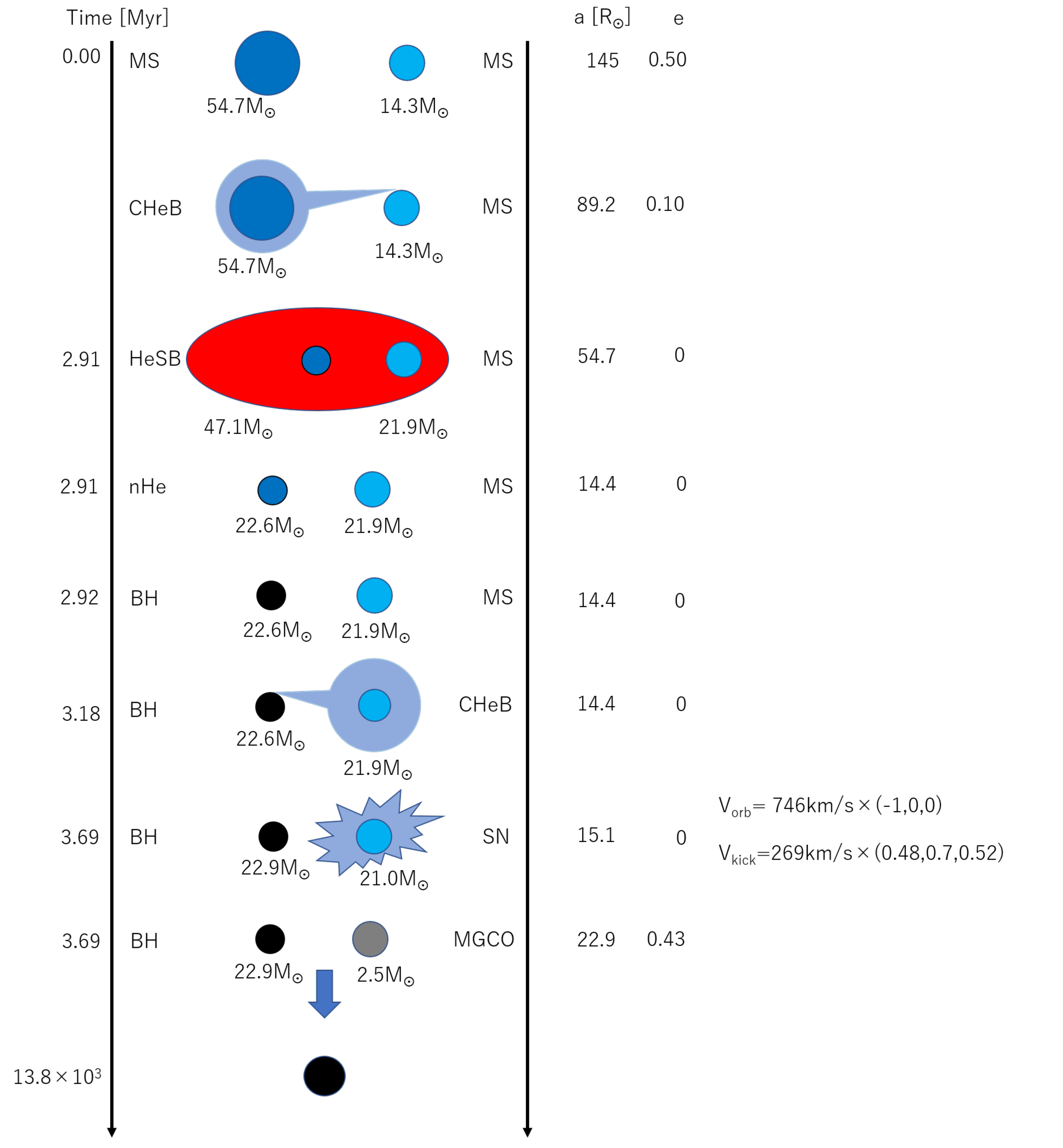}
    \caption{Evolutionary path of a Pop III BH-MGCO binary
    similar to GW190814. MS, CHeB, HeSB, and nHe mean the main sequence, Core He burning, He shell burning, and naked He stars, respectively.
    The naked He star is the remnant after the common envelope phase.
    $V_{\rm orb}$, and $V_{\rm kick}$ are the orbital velocity just before the SN, and the natal kick velocity, respectively.
    $a$ (in the solar radius $R_{\odot}$) and $e$ denote the orbital separation
    and eccentricity, respectively.}
    \label{fig:evolution_path}
\end{figure}

Since the typical chirp mass of BH-MGCO binary of $\sim 6M_{\odot}$ is almost the same as that of GW190814, we searched if similar BH-MGCO binary to GW190814 exists in our simulation data. As a result, we found one with mass $(22.9 M_{\odot},2.5M_{\odot})$. Figure~\ref{fig:evolution_path} shows the evolutionary path of this binary.
The binary is born as the zero age main sequence binary consisting of $54.7\,M_{\odot}$ and $14.3\,M_{\odot}$ stars.
The primary evolves to a giant and starts a mass transfer.
Next, the primary becomes so large that the secondary plunges into the primary envelope so that the binary enters common envelope phase.
In the common envelope phase, the primary envelope is evaporated and the separation shrinks. Without natal kick, the mass ejection at the SN of secondary makes the binary too wide to merge within the Hubble time.
However, if the direction of the natal kick is inverse to the orbital velocity, the binary is able to merge within the Hubble time. Figure.~\ref{fig:evolution_path} shows
that the final destiny of this binary is $22.9 M{\odot}$ BH and $2.5M{\odot}$ compact object which is either the lightest BH or the most massive NS. This example might be a possible answer to the question of ``What are the processes by which the lightest BHs
or the most massive NSs form in Ref.~\cite{Abbott:2020khf} ?'' in {\it Introduction}.

\begin{figure}[tb]
    \centering
    \includegraphics[width=0.8\textwidth]{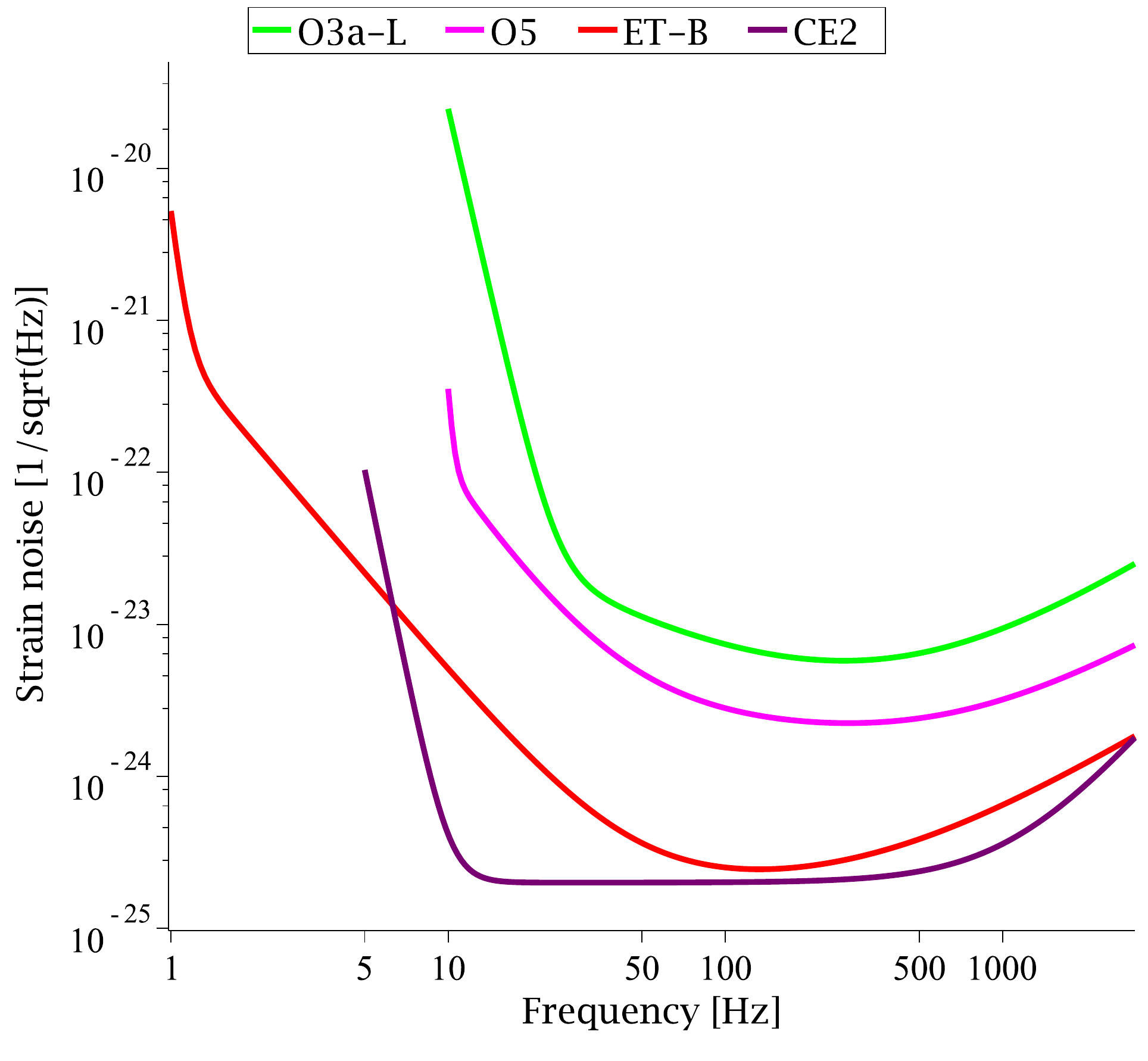}
    \caption{Strain-noise fitted curves
    for 4 GW detector configurations:
    LIGO O3a-Livingston (O3a-L, green),
    LIGO O5 (O5, magenta), Einstein Telescope (ET-B) (red)
    and Cosmic Explorer (CE2) (purple).
    These fitting curves are obtained by using 
    Refs.~\cite{Aasi:2013wya,Hild:2008ng,Mishra:2010tp,Reitze:2019iox}. We have lower frequency cutoffs of 10 Hz, 1 Hz
    and 5 Hz for O3a-L and O5,
    ET-B and CE2, respectively.}
    \label{fig:Strain_noise}
\end{figure}

To estimate the event rate of Pop III BH-MGCO binaries,
we use an inspiral-merger-ringdown waveform
presented in Ref.~\cite{Nakamura:2016hna}
which is based on Refs.~\cite{Dalal:2006qt,Ajith:2009bn},
The signal-to-noise ratio (SNR) of GW events
is calculated
in 4 strain-noise fitted curves
(see Fig.~\ref{fig:Strain_noise})
which are prepared by using Ref.~\cite{Aasi:2013wya}
for LIGO O3a-Livingston (O3a-L) (green) and LIGO O5 (magenta),
Ref.~\cite{Hild:2008ng} (see also Ref.~\cite{Mishra:2010tp})
for Einstein Telescope (ET-B) (red),
and Ref.~\cite{Reitze:2019iox}
for Cosmic Explorer (CE2) (purple).
Then, for example, the maximum observable redshift
$z_{\rm max}$
by setting the averaged SNR $=8$ for the GW190814 binary
with $m_1=23.2\,M_{\odot}$ and $m_2=2.59\,M_{\odot}$
becomes $z_{\rm max} = 0.0814$ for LIGO O3a-Livingston,
$0.211$ for LIGO O5, $5.81$ for ET-B, and $49.4$ for CE2.

Table~\ref{tab:ER} shows the averaged masses of BH-MGCO binaries in the solar mass $M_{\odot}$
for each kick model, and the event rates
in [${\rm yr}^{-1}$]
based on the maximum observable redshifts $z_{\rm max}$
(shown as values in parenthesis for each detector)
of this typical binaries for 4 GW detector configurations.
For the O3a-L detector, the maximum event rate of GWs is 
0.268 ${\rm yr}^{-1}$ for the kick model of
$\sigma_{\rm k}=150$ km/s.
For the O5, ET-B and CE2 detectors,
the maximum event rate of GWs are found as 
3.62 ${\rm yr}^{-1}$ for the $\sigma_{\rm k}=100$ km/s,
2070 ${\rm yr}^{-1}$ for the $\sigma_{\rm k}=265$ km/s
and 3830 ${\rm yr}^{-1}$ for the $\sigma_{\rm k}=300$ km/s,
respectively.

\begin{table}[!t]
\caption{
Averaged masses
($\langle m_1 \rangle$, $\langle m_2 \rangle$)
of BH-MGCO binaries in the solar mass $M_{\odot}$
and event rates in [${\rm yr}^{-1}$]
for 4 GW detector configurations:
LIGO O3a-Livingston (O3a-L), LIGO O5 (O5),
Einstein Telescope (ET-B) and Cosmic Explore (CE2)
in 8 kick models with
$\sigma_{\rm k}=0,~100,~150,~200,~265,~300,~400$,
and $500$ km/s.
Given the masses of binaries,
we estimate the maximum observable redshift
$z_{\rm max}$ (values in parenhtesis for each detector),
and then the event rates
for each detector are derived
by using the merger rate density calculated
by the population synthesis simulations of Pop III stars.
Although we can observe $z \sim 40$ by using the CE detector,
the merger rate density is approximately 0 
for $z \gtrsim 2.30$ in the cases of $\sigma_{\rm k}=0$,
for $z \gtrsim 32.75$ in the cases of $\sigma_{\rm k}=100$
and $150$ km/s, and for $z \gtrsim 33.30$
in the cases of $\sigma_{\rm k}=200,~265,~300,~400$
and $500$ km/s.}
\label{tab:ER}
\begin{center}
\begin{tabular}{cccccc}
\hline
$\sigma_{\rm k}$ (km/s)
& ($\langle m_1 \rangle$, $\langle m_2 \rangle$)
& O3a-L & O5 & ET-B & CE2 \\
\hline
0 & (36.9, 2.91) & 0.306 (0.106) & 4.03 (0.277) 
& 102 (6.39) &  102 (35.2) \\
100 & (32.1, 2.73) & 0.207 (0.0960) & 3.62 (0.250) 
& 687 (6.13) & 774 (38.7) \\
150 & (31.2, 2.67) & 0.268 (0.0938) & 3.47 (0.244) 
& 1440 (6.06) & 1850 (39.4) \\
200 & (30.9, 2.63) & 0.225 (0.0927) & 2.99 (0.241) 
& 2020 (6.01) & 3000 (39.6) \\
265 & (30.5, 2.59) & 0.194 (0.0914) & 2.92 (0.238) 
& 2070 (5.95) & 3750 (39.9) \\
300 & (30.3, 2.59) & 0.158 (0.0912) & 1.99 (0.237) 
& 1940 (5.95) & 3830 (40.1) \\
400 & (30.3, 2.59) & 0.111 (0.0912) & 1.13 (0.237) 
& 1560 (5.95) & 3430 (40.1) \\
500 & (30.4, 2.59) & 0.0928 (0.0913) & 1.24 (0.238) 
& 1200 (5.95) & 2700 (40.0) \\
\hline
\end{tabular}
\end{center}
\end{table}

\section{Discussion}

\begin{figure}[tb]
    \centering
    \includegraphics[width=0.8\textwidth]{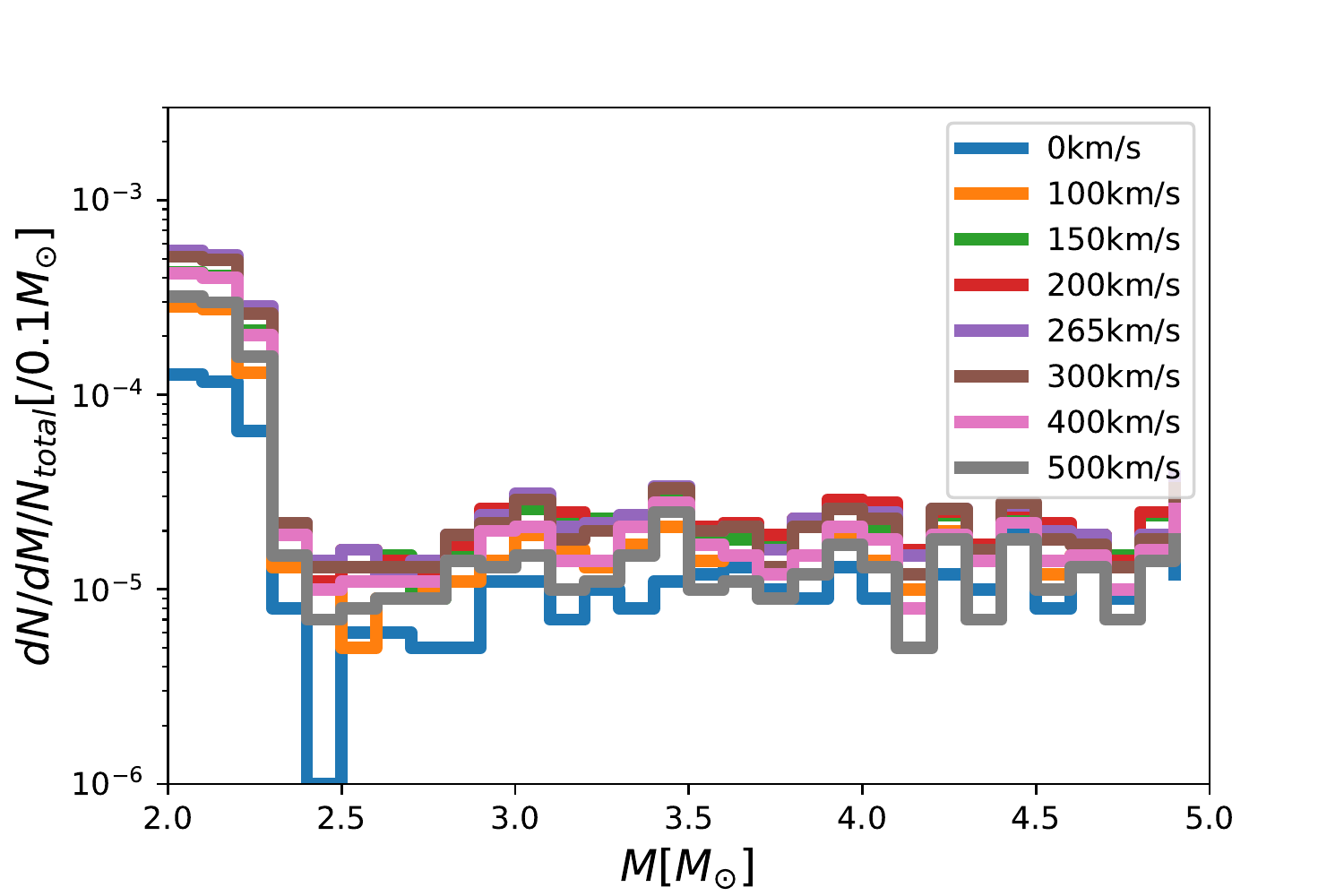}
    \caption{Mass distributions of MGCOs which merge within the Hubble time for 8 kick models with $\sigma_{\rm k}=0,~100,~150,~200,~265,~300,~400$, and $500$ km/s.
    We find that the lighter BHs can have various masses.}
    \label{fig:MGCOmass}
\end{figure}

In a binary system, a heavier BH is formed first.
After that, a lighter NS is formed.
At this time, a part of the blown outer layer
with a mass of $\sim 1\,M_{\odot}$ falls back,
and the NS becomes a BH with a mass of $\sim 2.6\,M_{\odot}$. Normally, this binary BH takes much longer time
to coalesce than the Hubble time,
but considering the kick velocity of the MGCO,
the binary coalescence will occur with in the Hubble time.
Figure \ref{fig:MGCOmass} shows mass distributions of MGCOs which merge within the Hubble time for each kick velocity model.
As a future prediction,
the lighter BHs can have various masses,
and some may be NS.
What seemed strange in the LIGO-Virgo paper~\cite{Abbott:2020khf}
is commonplace in this scenario.
The mass distribution of MGCOs (Fig. \ref{fig:MGCOmass}) does not depend on the kick velocity models, but depends on the SN remnant model.
Our understanding of the central engine of supernova
explosions
and also the study of the explosions
in Pop III are not yet complete.
Future detections of MGCOs may give a strong constraint on the SN remnant model.
As a summary, the lighter compact objects
will always become BH or NS
which is close to the maximum mass of NS
in the scenario of this letter.

\begin{figure}[tb]
    \centering
    \includegraphics[width=0.8\textwidth]{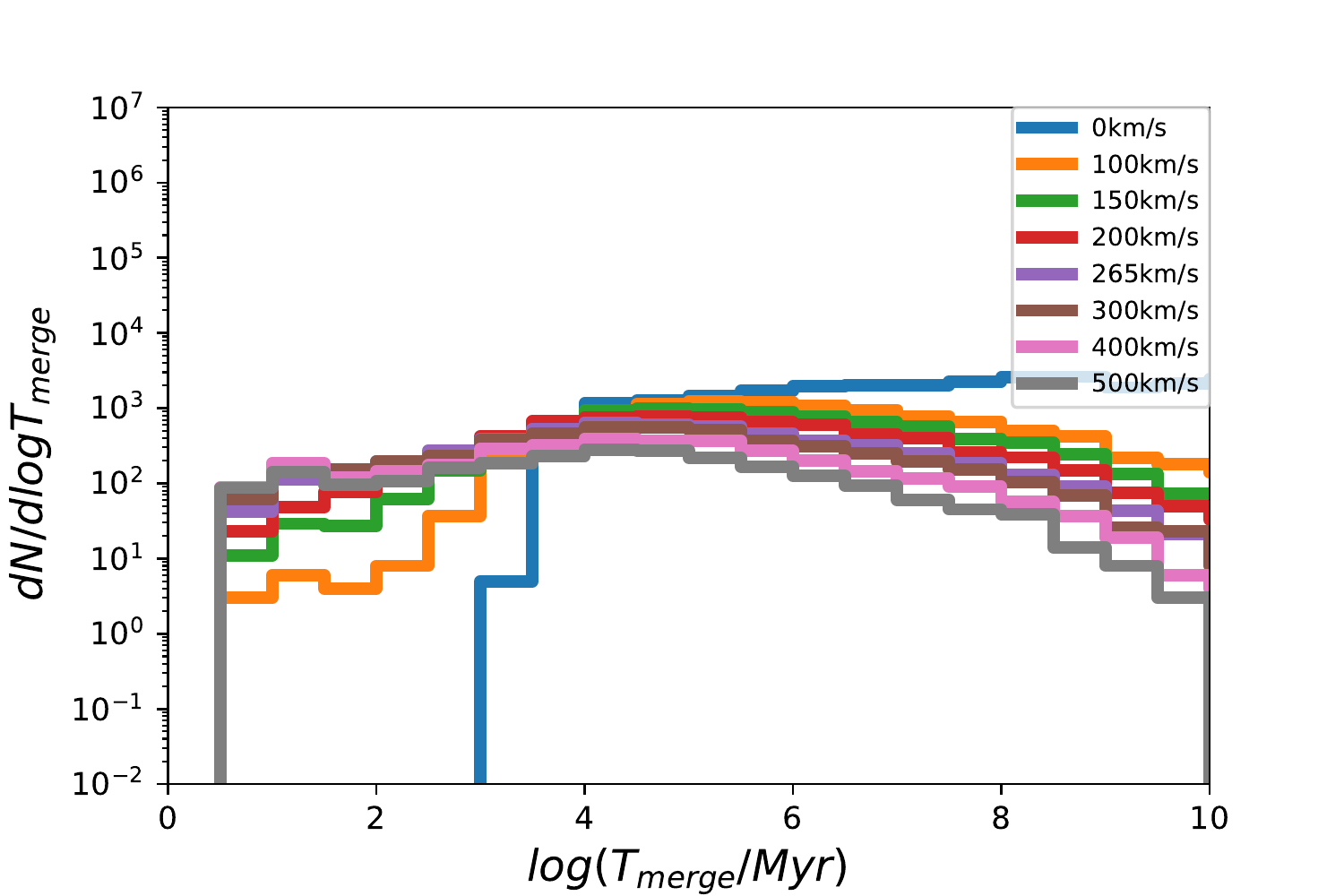}
    \caption{Merger time distributions for 8 kick models with with $\sigma_{\rm k}=0,~100,~150,~200,~265,~300,~400$, and $500$ km/s.}
    \label{fig:merger time}
\end{figure}

{In our simulation, the orbit of the binary before the supernova (SN) explosion of the secondary star after the BH formation of the primary star is eccentric in general. However in the eccentric orbit of the binary, it is very difficult to find the physical explanation of the effect of the kick velocity at the SN explosion using the analytic study so that we treat only a circular orbit before the SN explosion of the secondary star. We believe the essence of the physics can be obtained in our model.}

{Let us consider a binary with mass $m_1$ and $m_2~(< m_1)$  in a circular orbit with the separation $a$. 
We define the total mass by $m_t=m_1+m_2$. 
We take the center of the inertia at the origin of the $x$-$y$ coordinates so that the location of
the star 1 and the star 2 are at ${\bf r}_1=(-a_1,\,0)$ and ${\bf r}_2=(a_2,\,0)$ where $a_1=(m_2/m_t)\,a$ and $a_2=(m_1/m_t)\,a$,
respectively. From the results of Newtonian dynamics, the total Energy $E_0$, the angular momentum $J_0$ and
the angular frequency $\Omega$ of the binary are given by 
\begin{equation}
 E_0 = -\frac{G\,m_1m_2}{2a} \,, \quad
 J_0^2 = G\,m_1m_2\,\mu \,a \,, \quad
 \Omega = \sqrt{\frac{Gm_t}{a^3}} \,,
\label{eq:E_J_O}
\end{equation}
and
\begin{equation}
 \mu = \frac{m_1m_2}{m_1+m_2} \,,
\label{eq:mu}
\end{equation}
where $G$ is the gravitational constant.
The $y$-direction rotational velocity $v_1$ and $v_2$ are given by
\begin{equation}
 v_1 = -a_1\Omega \,, \quad v_2 = a_2\Omega \,.
\end{equation}}

{Now we regard that the primary star is a BH with mass $m_1=23\msun$ and the secondary one with mass $m_2=21\msun$ explodes as a supernova explosion remaining mass gap compact object with mass $m'_2=2.5\msun$ and a kick velocity  $v_k$ along the $y$-direction for simplicity.
Before the supernova, we assume that the orbit of the binary is expressed by Eqs.~\eqref{eq:E_J_O} and \eqref{eq:mu} with the location of
the BH at ${\bf r}_1$ and the pre-supernova secondary star at  ${\bf r}_2$. We treat the supernova as the instantaneous event so that after the supernova we regard that the secondary star turns out to be a mass gap compact object with mass $m'_2$ and only the $y$-direction velocity of $v'_2=v_2+v_k$ at ${\bf r}_2$.
The mass, location and velocity of the primary are not changed, that is, $m_1$, ${\bf r}_1$ and
$v'_1=v_1$. }

{The new position vector ${\bf r}_{\rm G}$ of the center of mass becomes
${\bf r}_{\rm G}=({\bf r}_1 m_1+{\bf r}_2 m'_2)/(m_1+m'_2)$ so that the $y$-component velocity of each star in the center of mass coordinate are 
\begin{equation}
 v''_1 = \frac{m'_2(v_1-v'_2)}{m_1+m'_2} \,, \quad
 v''_2 = \frac{-m_1(v_1-v'_2)}{m_1+m'_2} \,.
\label{eq:vdd}
\end{equation}
To obtain the semi-major axis $a_k$ and the eccentricity $e_k$ after the supernova explosion with the kick velocity $v_k$, let us calculate the energy $E_k$ and the angular momentum $J_k$ in the center of mass coordinate as
\begin{eqnarray}
 E_k&=&\frac{1}{2}m_1(v''_1)^2+\frac{1}{2}m'_2(v''_2)^2-\frac{Gm_1m'_2}{a} \,,
\label{eq:Ek}
\\
 J_k&=&(-a_1-x_G)\,m_1v''_1+(a_2-x_G)\,m'_2v''_2 \,,
\label{eq:Jk}
\end{eqnarray}
where
\begin{equation}
 x_G = \frac{-a_1m_1+a_2m'_2}{m_1+m'_2} \,.
\end{equation}
Inserting Eq.~\eqref{eq:vdd} into Eqs.~\eqref{eq:Ek} and \eqref{eq:Jk}, we have
\begin{eqnarray}
 E_k&=&\frac{1}{2}\,\mu' \left\{(a\Omega+v_k)^2-\frac{2G(m_1+m'_2)}{a} \right\} \,,
\label{eq:Ek2}
\\
 J_k&=&\mu'(a\Omega+v_k)\,a \,,
\label{eq:Jk2}
\end{eqnarray}
where
\begin{equation}
  \mu' = \frac{m_1m'_2}{m_1+m'_2} \,.
\label{eq:mud}
\end{equation}
Using $E_k$ and $J_k$, the new semi-major axis $a_k$ and eccentricity $e_k$ are given by
\begin{equation}
 a_k = -\frac{Gm_1m'_2}{2E_k} \,, \quad
 1-e_k^2 = -\frac{2E_k J_k^2}{G^2 m_1^2 {m'_2}^2 \mu'} \,.
\label{eq:ak_ek}
\end{equation}
Now the merging time $t_m$ due to the emission of the GWs is written as
\begin{equation}
 t_m=\frac{5}{256}\frac{a_k^4}{c}\left(\frac{Gm_1}{c^2}\right)^{-1}\left(\frac{Gm'_2}{c^2}\right)^{-1}\left(\frac{G(m_1+m'_2)}{c^2}\right)^{-1}(1-e_k^2)^{3.5} \,.
\label{eq:tm}
\end{equation}
Using Eqs.~\eqref{eq:Ek2} and \eqref{eq:Jk2}, Eq.~\eqref{eq:tm} is written as
\begin{equation}
 t_m=\frac{5}{256}\frac{1}{c}\left(\frac{Gm_1}{c^2}\right)^{-1}\left(\frac{Gm'_2}{c^2}\right)^{-1}\left(\frac{G(m_1+m'_2)}{c^2}\right)^{-1}\frac{J_k^7}{\sqrt{-2E_k}G^3 m_1^3 {m'_2}^2{\mu'}^{\frac{7}{2}}} \,.
\label{eq:tm2}
\end{equation}}

{Figure~\ref{fig:merger time} shows that if the kick velocity is zero, there are almost no mergers with a merging time less than $\sim 10^9~{\rm yr}$. This means that no GW signal is observed from the merger of binary BH-mass gap compact object around $z\sim 10$ since the age of the universe at $z\sim 10$ is $\sim4.5 \times 10^8~{\rm yr}$ which is much smaller than the merging time of the binary.
One of the methods to have  such mergers at $z\sim 10$ is to introduce the kick velocity in the SN explosion event of the secondary. If $(a\Omega+v_k)$ decreases, $J_k$ decreases and $\sqrt{-2E_K}$ increases from Eqs.~\eqref{eq:Ek2} and \eqref{eq:Jk2} so that the merging time decreases from Eq.~\eqref{eq:tm2}. 
Note here that to decrease $(a\Omega+v_k)$ under the same $a,\,m_1,\,m_2$ and $m'$, $v_k$ should be negative, that is, the direction of the kick is opposite to that of the rotation. The event rate of such mergers is roughly in proportion to the star formation rate such as de Souza et al.~(2011) star formation model of Pop III stars (see Fig.~8 of Kinugawa et al.~(2014)) adopted in our population synthesis simulation so that the formation of the binary starts from $z\sim 30$, has a peak at $z \sim 10$ and ends at $z\sim 5$. The merging time due to the emissions of GWs can be much smaller than the age of the universe for a given redshift $z$ so that for a given kick velocity, the event rate as a function of the redshift is found to be
similar to the star formation history in our population synthesis simulations. This theoretically expected behaviour of the merger rate as a function of $z$ agrees with Fig.~\ref{fig:merger rate}}

{For a fixed redshift $z$, our numerical simulation shows that the merging rate as a function of the kick velocity first increases, has a peak for $v_k\sim 300~{\rm km/s}$ and decreases again for $v_k >300~{\rm km/s}$. Since the merger rate for $v_k=0$ is almost zero, it is natural that the merging rate first increases as a function of $v_k$. While in the extreme limit of the large kick velocity, almost all BH-mass gap compact object systems disrupt so that the above increase of the merging rate should be stopped at some value of $v_k$, and after that it should decrease as a function of  $v_k$. In our case, we obtain the peak is $ v_k\sim 300~{\rm km/s}$ from numerical simulation. Although it is difficult to get the peak value of the velocity dispersion $\sigma_k$ analytically even in our circular orbit treatment of the binaries before SNe, we can confirm from the above physical arguments, the existence of the peak of the merger rate as a function of $v_k$ for a given redshift $z$. To obtain the numerical value of $v_k$, the population synthesis calculations are indispensable since the analytic arguments are restricted to the initial circular orbit and the direction of the kick velocity parallel to the orbital velocity.  }

{Using Eqs.~\eqref{eq:Ek2}--\eqref{eq:ak_ek}, we can rewrite Eq.~\eqref{eq:tm} as
\begin{equation}
\frac{a^4\left(1+\frac{v_k}{a\Omega}\right)^7}{\sqrt{2\left(\frac{m_1+m'_2}{m_1+m_2}\right)-\left(1+\frac{v_k}{a\Omega}\right)^2}} =\frac{256}{5}\left(\frac{Gm_1}{c^2}\right)\left(\frac{Gm'_2}{c^2}\right)\left(\frac{G(m_1+m'_2)}{c^2}\right)\left(\frac{m_1+m'_2}{m_1+m_2}\right)^3\,c\,t_m \,.
\end{equation}
Inserting the adopted values of $m_1=23\msun$, $m_2=21\msun$ and $m'_2=2.5\msun$,
we have
\begin{equation}
\frac{a^4\left(1+\frac{v_k}{a\Omega}\right)^7}{\sqrt{1.159-\left(1+\frac{v_k}{a\Omega}\right)^2}}
=(8.26\times 10^{11}{\rm cm})^4\left(\frac{t_m}{10^{10}{\rm yr}}\right) \,.
\label{eq:case}
\end{equation}}

{As for the merger rate at $z=0$, the merger time of the binary should be $t_m\sim10^{10}$y so that even in the case of the zero kick velocity, some binaries
merge at $z=0$. Now for $t_m=10^{10}$~yr, we can rewrite Eq.~\eqref{eq:case} as
\begin{equation}
    a\sim 6.57\times 10^{11}~{\rm cm}~\frac{\left\{1-\frac{2}{0.159}\left(\frac{v_k}{a\Omega}\right)- \frac{1}{0.159}\left(\frac{v_k}{a\Omega}\right)^2 \right\}^{\frac{1}{8}}}{\left(1+\frac{v_k}{a\Omega}\right)^\frac{7}{4}} \,.
\label{eq:a(vk)_z=0}
\end{equation}
For $v_k= 0$, $a\sim 6.57\times 10^{11}~{\rm cm}$.
While for $v_k\neq 0$ the binary with $a$ defined by
Eq.~\eqref{eq:a(vk)_z=0} merges at $z=0$. For a given Gaussian kick velocity distribution with a velocity dispersion of $\sigma_k$, and for various $v_k\neq 0$ with $a$ determined by Eq.~\eqref{eq:a(vk)_z=0}, the merger time can be $t_m=10^{10}$~yr so that the number of mergers should increase compared with that for $v_k=0$. However in the limit of extreme large $\sigma_k$, the merger rate should decrease again
since the most of $v_k$ in an extremely large $\sigma_k$ can not satisfy Eq.~\eqref{eq:a(vk)_z=0}.
Therefore, we can confirm that the event rate as a function of  $\sigma_k$, first increases, has a peak, and then decreases from this physical argument.
So far, we adopted only the circular orbit initially. In reality, the circular orbit approximation of binaries is not the case in general so that it is hard to estimate the peak value of $\sigma_k$  and the merger rate analytically. However from the above physical arguments we can confirm the existence of the peak of the event rate as a function of $\sigma_k$ definitely. To estimate them, we need the population synthesis numerical simulations as we did in our paper.}

\section*{Acknowledgment}

~~~T. K. acknowledges support from University of Tokyo Young Excellent researcher program.
T. N. acknowledges support from
JSPS KAKENHI Grant No. JP15H02087.
H. N. acknowledges support from
JSPS KAKENHI Grant Nos. JP16K05347 and JP17H06358.



\begin{thebibliography}{DUM}

\bibitem{Abbott:2020khf}
R.~Abbott \textit{et al.} [LIGO Scientific and Virgo],
Astrophys. J. \textbf{896}, L44 (2020)
[arXiv:2006.12611 [astro-ph.HE]].

\bibitem{glossary}
\url{https://emfollow.docs.ligo.org/userguide/glossary.html}


\bibitem{Rastello:2020sru}
S.~Rastello, M.~Mapelli, U.~N.~Di Carlo, N.~Giacobbo, F.~Santoliquido, M.~Spera, A.~Ballone and G.~Iorio,
[arXiv:2003.02277 [astro-ph.HE]].

\bibitem{Broadhurst:2020cvm}
T.~Broadhurst, J.~M.~Diego and G.~F.~Smoot,
[arXiv:2006.13219 [astro-ph.CO]].

\bibitem{Zevin:2020gma}
M.~Zevin, M.~Spera, C.~P.~L.~Berry and V.~Kalogera,
[arXiv:2006.14573 [astro-ph.HE]].

\bibitem{Most:2020bba}
E.~R.~Most, L.~J.~Papenfort, L.~R.~Weih and L.~Rezzolla,
[arXiv:2006.14601 [astro-ph.HE]].

\bibitem{Vattis:2020iuz}
K.~Vattis, I.~S.~Goldstein and S.~M.~Koushiappas,
[arXiv:2006.15675 [astro-ph.HE]].

\bibitem{Tan:2020ics}
H.~Tan, J.~Noronha-Hostler and N.~Yunes,
[arXiv:2006.16296 [astro-ph.HE]].

\bibitem{Safarzadeh:2020ntc}
M.~Safarzadeh and A.~Loeb,
[arXiv:2007.00847 [astro-ph.HE]].

\bibitem{Essick:2020ghc}
R.~Essick and P.~Landry,
[arXiv:2007.01372 [astro-ph.HE]].

\bibitem{Zhang:2020zsc}
N.~B.~Zhang and B.~A.~Li,
[arXiv:2007.02513 [astro-ph.HE]].

\bibitem{Jedamzik:2020omx}
K.~Jedamzik,
[arXiv:2007.03565 [astro-ph.CO]].

\bibitem{Tsai:2020hpi}
Y.~D.~Tsai, A.~Palmese, S.~Profumo and T.~Jeltema,
[arXiv:2007.03686 [astro-ph.HE]].

\bibitem{Fattoyev:2020cws}
F.~J.~Fattoyev, C.~J.~Horowitz, J.~Piekarewicz and B.~Reed,
[arXiv:2007.03799 [nucl-th]].

\bibitem{Mandel:2020cig}
I.~Mandel, B.~Mueller, J.~Riley, S.~E.~de Mink, A.~Vigna-Gomez and D.~Chattopadhyay,
[arXiv:2007.03890 [astro-ph.HE]].

\bibitem{Yang:2020xyi}
Y.~Yang, V.~Gayathri, I.~Bartos, Z.~Haiman, M.~Safarzadeh and H.~Tagawa,
[arXiv:2007.04781 [astro-ph.HE]].

\bibitem{Tsokaros:2020hli}
A.~Tsokaros, M.~Ruiz and S.~L.~Shapiro,
[arXiv:2007.05526 [astro-ph.HE]].

\bibitem{Tews:2020ylw}
I.~Tews, P.~T.~H.~Pang, T.~Dietrich, M.~W.~Coughlin, S.~Antier, M.~Bulla, J.~Heinzel and L.~Issa,
[arXiv:2007.06057 [astro-ph.HE]].

\bibitem{Clesse:2020ghq}
S.~Clesse and J.~Garcia-Bellido,
[arXiv:2007.06481 [astro-ph.CO]].

\bibitem{Lim:2020zvx}
Y.~Lim, A.~Bhattacharya, J.~W.~Holt and D.~Pati,
[arXiv:2007.06526 [nucl-th]].

\bibitem{Dexheimer:2020rlp}
V.~Dexheimer, R.~O.~Gomes, T.~Klähn, S.~Han and M.~Salinas,
[arXiv:2007.08493 [astro-ph.HE]].

\bibitem{Sedrakian:2020kbi}
A.~Sedrakian, F.~Weber and J.~J.~LI,
[arXiv:2007.09683 [astro-ph.HE]].

\bibitem{Roupas:2020jyv}
Z.~Roupas,
[arXiv:2007.10679 [gr-qc]].

\bibitem{Godzieba:2020tjn}
D.~A.~Godzieba, D.~Radice and S.~Bernuzzi,
[arXiv:2007.10999 [astro-ph.HE]].


\bibitem{Kinugawa:2016skw}
T.~Kinugawa, T.~Nakamura and H.~Nakano,
PTEP \textbf{2017}, 021E01 (2017)
[arXiv:1610.00305 [astro-ph.HE]]. 

\bibitem{Kinugawa:2014zha}
T.~Kinugawa, K.~Inayoshi, K.~Hotokezaka, D.~Nakauchi and T.~Nakamura,
Mon. Not. Roy. Astron. Soc. \textbf{442}, 2963-2992 (2014)
[arXiv:1402.6672 [astro-ph.HE]].

\bibitem{Belczynski:2001uc}
K.~Belczynski, V.~Kalogera and T.~Bulik,
Astrophys. J. \textbf{572}, 407-431 (2001)
[arXiv:astro-ph/0111452 [astro-ph]].

\bibitem{Kochanek2020}
C. S.~Kochanek
Mon. Not. Roy. Astron. Soc. \textbf{493} 4945 (2020)
[arXiv:2001.07216 [astro-ph.SR]].




\bibitem{Nakamura:2016hna}
T.~Nakamura, M.~Ando, T.~Kinugawa, H.~Nakano, K.~Eda, S.~Sato, M.~Musha, T.~Akutsu, T.~Tanaka, N.~Seto, N.~Kanda and Y.~Itoh,
PTEP \textbf{2016}, 093E01 (2016)
[arXiv:1607.00897 [astro-ph.HE]].

\bibitem{Ajith:2009bn}
P.~Ajith, M.~Hannam, S.~Husa, Y.~Chen, B.~Bruegmann, N.~Dorband, D.~Muller, F.~Ohme, D.~Pollney, C.~Reisswig, L.~Santamaria and J.~Seiler,
Phys. Rev. Lett. \textbf{106}, 241101 (2011)
[arXiv:0909.2867 [gr-qc]].

\bibitem{Dalal:2006qt}
N.~Dalal, D.~E.~Holz, S.~A.~Hughes and B.~Jain,
Phys. Rev. D \textbf{74}, 063006 (2006)
[arXiv:astro-ph/0601275 [astro-ph]].

\bibitem{Aasi:2013wya}
B.~P.~Abbott \textit{et al.} [KAGRA, LIGO Scientific and VIRGO],
Living Rev. Rel. \textbf{21}, 3 (2018)
[arXiv:1304.0670 [gr-qc]].

\bibitem{Hild:2008ng}
S.~Hild, S.~Chelkowski and A.~Freise,
[arXiv:0810.0604 [gr-qc]].

\bibitem{Mishra:2010tp}
C.~K.~Mishra, K.~G.~Arun, B.~R.~Iyer and B.~S.~Sathyaprakash,
Phys. Rev. D \textbf{82}, 064010 (2010)
[arXiv:1005.0304 [gr-qc]].

\bibitem{Reitze:2019iox}
D.~Reitze, R.~X.~Adhikari, S.~Ballmer, B.~Barish, L.~Barsotti, G.~Billingsley, D.~A.~Brown, Y.~Chen, D.~Coyne, R.~Eisenstein, M.~Evans, P.~Fritschel, E.~D.~Hall, A.~Lazzarini, G.~Lovelace, J.~Read, B.~S.~Sathyaprakash, D.~Shoemaker, J.~Smith, C.~Torrie, S.~Vitale, R.~Weiss, C.~Wipf and M.~Zucker,
Bull. Am. Astron. Soc. \textbf{51}, 035
[arXiv:1907.04833 [astro-ph.IM]].



\end{thebibliography}
\end{document}